\documentclass[letterpaper,twocolumn,prl,showpacs,floatfix,
groupedaddress]{revtex4}
\usepackage{graphicx}

\begin{document}

\title{Phase diagram of the dilute magnet LiHo$_x$Y$_{1-x}$F$_4$}

\author{A. Biltmo}
\author{P. Henelius}
\affiliation{Dept. of Theoretical Physics, Royal Institute of Technology,
SE-106 91 Stockholm, Sweden}

\date{\today}

\begin{abstract}

  We study the effective long-range Ising dipole model with a local
  exchange interaction appropriate for the dilute magnetic compound
  LiHo$_{x}$Y$_{1-x}$F$_4$. Our calculations yield a value of 0.12 K for
  the nearest neighbor exchange interaction. Using a Monte Carlo
  method we calculate the phase boundary $T_c(x)$ between the
  ferromagnetic and paramagnetic phases. We demonstrate that the
  experimentally observed linear decrease in $T_c$ with dilution is
  not the simple mean-field result, but a combination of the effects
  of fluctuations, the exchange interaction and the hyperfine
  coupling.  Furthermore, we find a critical dilution $x_c=0.21(2)$,
  below which there is no ordering. In agreement with recent Monte
  Carlo simulations on a similar model, we find no evidence of the
  experimentally observed freezing of the glassy state in our
  calculation.  We apply the theory of Stephen and Aharony to
  LiHo$_{x}$Y$_{1-x}$F$_4$ and find that the theory does predict a
  finite-temperature freezing of the spin glass. Reasons for the
  discrepancies are discussed.
\end{abstract}

\maketitle

%\section{Introduction}
The rare-earth compound LiHo$_x$Y$_{1-x}$F$_4$ has been widely used as
a model magnet displaying a wide range of phenomena.  At T$_c$=1.53 K
the predominant long-range dipolar interaction causes a second order
classical phase transition to a ferromagnetic state\cite{rose91}. By
applying a transverse magnetic field the order can be destroyed in a
T=0 quantum phase transition at about 4.9 T\cite{rose96}. Positional
disorder can be introduced by substituting the magnetic Ho$^{3+}$ ions
with non-magnetic Y$^{3+}$ ions. The disorder has been shown to cause
a transition to glassy behavior at high dilution\cite{rose90}. 
%In the
%limit of extreme dilution the glass appears not to freeze, while at
%high dilution spin glass freezing was observed.

A main attraction of LiHo$_x$Y$_{1-x}$F$_4$ is that the microscopic
model is well-known\cite{rose90, chak04}. The ground state of the
Ho$^{3+}$ ion in the crystal field is an Ising doublet, with the first
excited state 11 K above the ground state. At the temperature range we
consider here (T $< 1.5$ K) LiHoF$_{4}$ should be a very good realization of
a dipolar Ising model
\begin{equation}
H=\frac{J}{2}\sum_{i\ne j}\frac{r_{ij}^2-3z_{ij}^2}{r_{ij}^5}
\sigma_i^z\sigma_j^z+ \frac{J_{\rm{ex}}}{2}\sum_{i,nn}\sigma^{z}_{i}\sigma^{z}_{nn}
 \label{dipole}
\end{equation}
where $J$ is the dipolar coupling constant, $ J_{\rm{ex}}$ the
nearest-neighbor exchange constant, $r_{ij}$ the interspin distance
and $z_{ij}$ the interspin distance along the Ising axis. The
summation is done over all Ho$^{3+}$ ions, which form a tetragonal
Bravais lattice with four ions per unit cell.  When diluted, a
fraction $x$ of the sites are occupied by non-magnetic Yttrium and not
included in the above sum. The size of the unit cell is $(1,1,2.077)$
in units of $a=5.175 \mathring{A}$. If we express the interspin
distance in units of $a$, then the dipolar coupling constant
$J=(g\mu_B/2)^2/a^3=0.214 K $\cite{chak04}. The exchange coupling
$J_{\rm{ex}}$ has been experimentally determined to about half of the
nearest-neighbor dipolar coupling\cite{menn84}. In our calculation we
have neglected the next nearest neighbor exchange interaction, which was found
to be about $5\%$ of the nearest-neighbor dipolar
coupling\cite{menn84}. In addition, we have left out the hyperfine
coupling between the nuclear and electronic spins as well as the
random fields generated by the breaking of crystal symmetries due to
the dilution. The effects of these terms on our results will be
discussed.

A goal of the extensive experimental studies\cite{rose90} of the
dilute magnet LiHo$_x$Y$_{1-x}$F$_4$ is to establish the material as a
spin glass prototype with canonical glass properties, and with a well
understood microscopic theory. This would allow comparison between
different analytical approaches to spin-glass systems, as well as
provide an important experimental benchmark. Currently, it is widely
believed that the above dipolar Ising model captures the essential
behavior of LiHo$_x$Y$_{1-x}$F$_4$ observed in numerous experiments,
yet a direct calculation of the phase diagram is lacking. The goal of
this study is to fill this void and determine the phase diagram for
the dilute dipolar Ising model appropriate for LiHo$_x$Y$_{1-x}$ by a
direct non-approximate Monte Carlo calculation. In the process we also
address the fundamental question of whether a disordered classical
dipolar ferromagnet supports a long-ranged spin-glass phase.

\begin{figure}[htp]
\resizebox{\hsize}{!}{\includegraphics[clip=true]{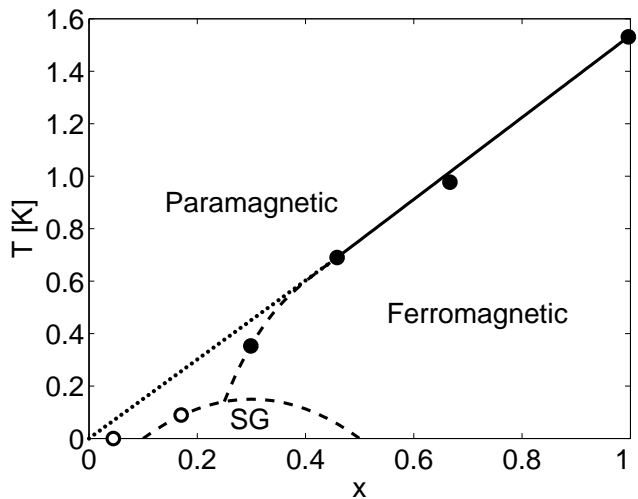}}
\caption{\label{expPhDiag} Experimental phase diagram from Ref.~\onlinecite{rose90}. Open circles denote glassy behaviour, SG = spin glass.}
\end{figure}

The experimentally obtained phase diagram in shown in
Fig.~\ref{expPhDiag}. For $x>0.5$ the boundary between the
paramagnetic and ferromagnetic phases can be fitted to a straight line
passing through the origin, corresponding to the mean-field result
T$_c$(x)=xT$_c$(1). As the dilution is increased the boundary falls
below the mean-field result and glassy behavior ensues. At one point
(x=0.167) freezing of the spin glass was observed and at further
dilution (x=0.045) the glassy state did not appear to freeze. This
so-called anti-glass phase shows a behavior distinct from traditional
spin glasses and has been the subject of numerous
investigations\cite{rose87,ghos02, quil06}.

We are aware of two earlier theoretical investigations of randomly
parked dipoles. The conclusion of the first study\cite{step81},
considering bond-diluted dipoles, was that, depending on the lattice
structure, spin-glass ordering may be favored over ferromagnetic
ordering at low-T. The ordering (spin glass or ferromagnetic) persists
for any finite dilution $x$, in disagreement with the anti-glass
phase.  The second study\cite{xu91} predicts that a site-diluted BCC
lattice is ferromagnetically ordered above x=0.21 with a spin-glass
phase below x=0.21. It is also interesting to note that a study of the
three dimensional RKKY Ising spin glass, with an interaction of mixed
sign proportional to $1/r^3$, finds that this system lies on the
boundary between a finite temperature and a $T_c=0$ spin
glass\cite{bray86}.

Numerical Monte Carlo studies of dipoles on a dilute BCC
lattice\cite{xu91} find a transition to ferromagnetic ordering at
$x=0.3\pm 0.1$, but are unable to determine whether there is a low-T
spin glass transition. A more recent Monte Carlo study of Ising
dipoles\cite{snid05} on a cubic lattice at dilutions x=0.045, 0.12 and
0.20 fails to find a finite-temperature spin-glass transition. Note
that the dipolar model on a cubic lattice is not a ferromagnet at
higher temperatures, unlike LiHoF$_4$. In conclusion, the most
relevant theoretical and numerical studies to date disagree with
experiments on the existence and extent of the glassy low-T part of
the phase diagram. This could be partially explained by the subtleties
of the dipolar interaction since numerical and theoretical predictions
depend on the lattice structure and boundary conditions
used\cite{lutt46, kret79}. Our goal is therefore to tailor our
calculations to LiHo$_x$Y$_{1-x}$F$_4$ in order to be able to compare
the entire phase diagram with experiments.

%\section{Computational methods}
We have studied the dipolar Ising model given by Eq.~(\ref{dipole})
using a Monte Carlo method. Due to the long-range nature and angular
dependence of the Hamiltonian this is a challenging problem. Luttinger
and Tisza\cite{lutt46} demonstrated that lattice sums depended on
the sample shape, while Griffiths later showed\cite{grif68} that
physical properties are independent of sample shape due to break-up
into sample-shape dependent domains. In LiHoF$_{4}$ there is clear
experimental evidence for long needle-shaped domains\cite{cook75,
  batt75}. In order to compare calculations to experiments the domain
structure has to be taken into account, and there are, at present, two
different approaches\cite{kret79}. Previously the domain structure of
LiHoF$_{4}$ was taken into account by performing the Monte Carlo simulation
over a spherical cavity embedded in a cylindrical domain\cite{chak04}.
The part of the domain external to the cavity is treated in mean-field
theory and gives rise to an effective field acting on the sphere.

Here we choose the other approach, which is to impose periodic
boundary conditions and evaluate the effective interaction between
spin $i$ and $j$ as a sum over all periodic images of spin j. It is
important that the thermodynamic limit reflects the domain shape. For
a needle shaped domain, which is relevant for LiHoF$_{4}$, this means
carrying out the sum along the Ising axis prior to the sum in the
radial direction.  A significant speed-up in evaluation the sums can
be achieved using the Ewald summation method, which splits the sum
into two rapidly converging parts, one in Fourier space, and one in
real space. The advantages with periodic boundary conditions over the
cavity method are twofold. The cavity method neglects all fluctuations
outside the spherical cavity while the periodic images include at
least part of the fluctuations in the domain. The cavity method was
also shown to lead to non-monotonic system-size dependence in some
quantities\cite{chak04}, which is not the case for periodic boundary
conditions.
 
Due to the long-range interactions, the time required for one Monte
Carlo step scales as $N^2$, as opposed to $N$ for the short-range
case. Adding the computational expense of performing disorder averages
over several hundred copies of the system makes the efficiency of the
Monte Carlo method particularly important. We have therefore compared
the efficiency of the single spin-flip Metropolis method with
continuous time Monte Carlo\cite{bort75}, the SSE cluster
algorithm\cite{chak04} and the Wang-Landau
method\cite{2001PhRvL..86.2050W}, which gives explicit access to the
density of states. In agreement with other studies we found that the
Wang-Landau method converges very slowly for large system sizes. The
cluster algorithm allows for inclusion of a transverse field, but in
the present low-temperature classical simulations it becomes
inefficient since all spins tend to join a single cluster. The
continuous time Monte Carlo method also proved less efficient than the
traditional single-spin flip, which therefore was used throughout this
study.
 
%\section{Results}
In order to determine the extent of the ferromagnetic phase, the
critical temperature $T_c$ is determined as a function of disorder
$x$. In the Monte Carlo simulation this is accomplished by calculating
the Binder ratio for the magnetization
\begin{equation}
g_m= \left\langle 1-\frac{\langle M^4 \rangle}{3 \langle M^2 \rangle^2 }\right\rangle_d.
\end{equation}
In addition to the thermal average, an average over
quenched disorder configurations $d$ is calculated. The critical
temperature was extracted from the intersection of the Binder ratio
for different system sizes. We used system sizes up to $10^3$ unit cells,
containing 4000 spins. Disorder averages were performed over a few
hundred disorder configurations. A typical run consisted of
$2\times10^6$ Monte Carlo steps of which the first $10^6$ steps were
discarded.

In mean-field theory there are two phases, a low-temperature
ferromagnetic phase and a high-temperature paramagnetic phase
separated by a phase boundary $T_c(x)=xT_c(1)$. For the present model
$T_c(1)=2.41$ K in simple mean-field theory\cite{chak04},
significantly higher than the experimental value of 1.53 K. The
effects of fluctuations can be included using a Monte Carlo method,
and a recent study using the cavity method found that $T_c(1)=2.03$
K\cite{chak04}. In the present study the periodic boundary conditions
allow for fluctuations in the domain surrounding the Monte Carlo cell,
and we find that $T_c(1)=1.91$ K for the clean system. The difference
between the present and the experimental result can be attributed to
an anti-ferromagnetic exchange interaction which was measured to about
half of the nearest neighbor dipolar interaction\cite{menn84}.
Treating $J_{\rm{ex}}$ as a free parameter we find that a value of
$J_{\rm{ex}}=0.12$ K, or about 38 $ \% $ of the nearest neighbor
dipolar interaction $J_{\rm{dip}}^1=0.33 $ K, lowers $T_c$ to 1.53 K.

\begin{figure}[htp]
\resizebox{\hsize}{!}{\includegraphics[clip=true]{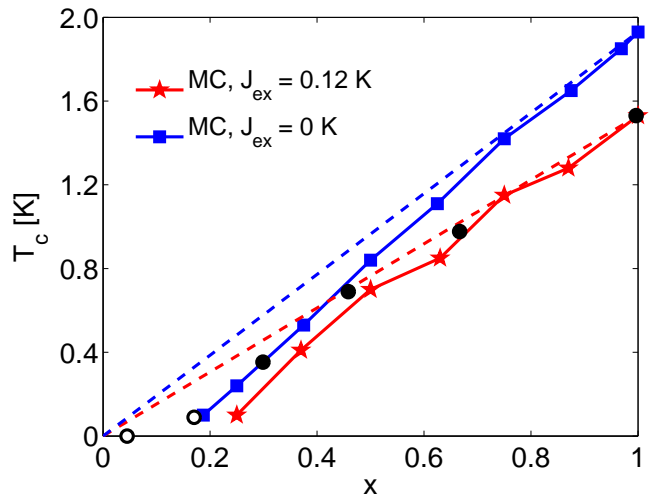}}
\caption{\label{Tc} $T_c$ as a function of dilution from experiments
  (circles) and Monte Carlo calculations. The dashed lines represent
  mean-field solutions.}
\end{figure}

In Fig.~\ref{Tc} We display the $T_c(x)$ boundary for Monte Carlo
and mean-field theory and compare it to the experimental data from
Ref.~\onlinecite{rose90}.

At low and intermediate dilution, $(x <0.5)$, the three experimental
data points follow the mean-field solution. In the Monte Carlo data
the effects of fluctuations are visible already around $x=0.7$,
particularly without exchange. Including the exchange term makes this
effect less visible and the Monte Carlo data is in quite good
agreement with experiments down to x=0.5. However, the Monte Carlo
data do fall increasingly below the experimental results as the
dilution is increased. One reason for this small difference is
probably the hyperfine coupling between the nuclear and electronic
spins\cite{rose96, chak04}. This term is important in the
low-temperature regime and omitted in our analysis. The general effect
of the hyperfine coupling is to increase the order, and its omission
would explain why $T_c(x)$ decreases faster with higher dilution for
the Monte Carlo data than for the experimental data. We have therefore
demonstrated that the experimentally observed linear decrease in $T_c$
is not the simple mean-field result, but rather a combination of the
effects of fluctuations, the exchange interaction and the hyperfine
coupling.

In agreement with the experimental data our phase boundary appears to
intersect the x-axis at a finite value of the dilution. This is in
sharp contrast to theoretical studies\cite{xu91, step81} that predict
a phase boundary extending to the origin. Extrapolating our data the
phase boundary intersects the x-axis at about $x_c=0.15(2)$ (no
exchange), and at $x_c$=0.21(2) (including exchange). This is close to
$x=0.167$, where experiments observed freezing of a spin glass at
$T_c=0.13$ K. In order to find signs of a spin glass freezing we have
performed independent simulations of two replicas (same quenched
disorder) simultaneously and the Edwards-Anderson overlap,
\begin{eqnarray}
q= \sum_i \sigma^{(1)}_i \sigma^{(2)}_i,
\end{eqnarray}
has been recorded. For a spin glass freezing to occur there should be
an intersection of the overlap Binder cumulants, $g_q$, but no
intersection of the magnetic Binder cumulant, $g_m$.

\begin{figure}[htp]
\resizebox{\hsize}{!}{\includegraphics[clip=true]{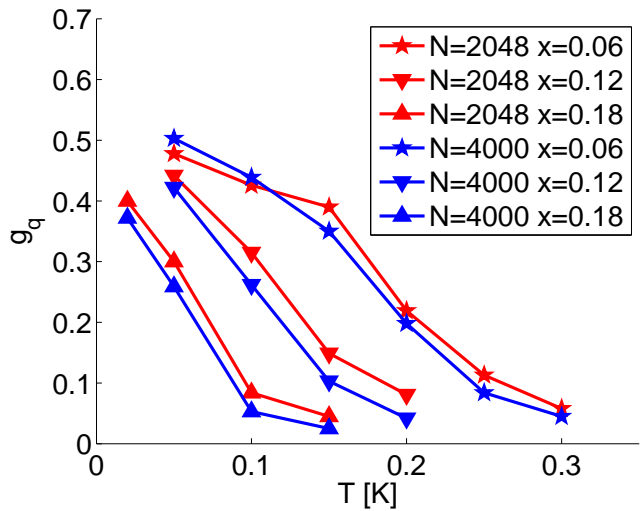}}
\caption{\label{overlapBC} Overlap Binder cumulants in the limit of
  high dilution.}
\end{figure}

We show the results for the overlap cumulant in
Fig.~\ref{overlapBC}. The data shown is for the case of no exchange
interaction, but we found similar results when including the exchange
term.  For $x=0.18$ the curves intersect around $T=0.12$ K, but the
magnetic Binder cumulant also intersects at this point, and we conclude
that the system is magnetized. When we increase the dilution the
curves do not intersect and we conclude that there is no finite
temperature freezing of the spin glass above $T=0.05$ K. At temperatures
lower than $T=0.05$ K equilibration problems occur and we cannot exclude
the possibility of freezing. However, the experimentally observed
freezing for $x=0.17$ occurred at $T=0.13$ K, and should be visible in
our data.

\begin{figure}[htp]
\resizebox{\hsize}{!}{\includegraphics[clip=true]{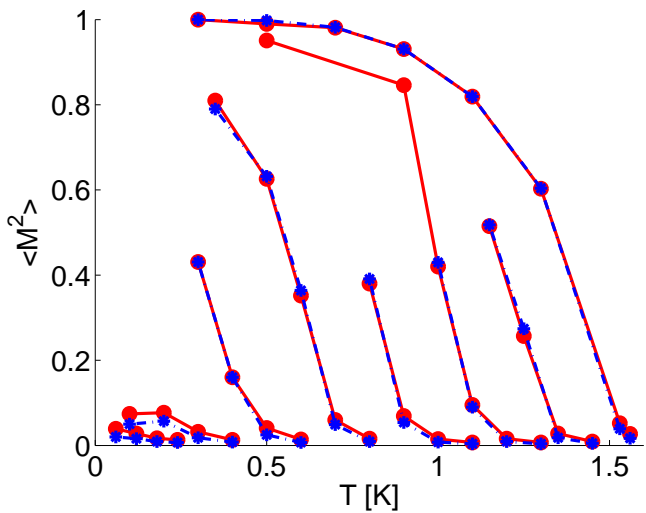}}
\caption{\label{Msquares} Magnetization squared  for $x=n/32$ with
  $n=4, 8, 12, 16, 20, 24, 28$ and 32 (left to right) for N=4000
  (dashed line) and N=2048 (solid line).}
\end{figure}

In order to give further credibility to the phase diagram in
Fig.~\ref{Tc} we plot the magnetization squared as a function of
disorder in Fig.~\ref{Msquares}. We note that except for the two
most diluted systems the finite-size effects are very small for the
system sizes considered (N=4000 and 2048). In the limit of high
dilution the magnetization decreases with increasing system size,
indicative of the lack of magnetic order.

In order to compare our results to theory we have applied the
mean-field calculation of Stephen and Aharony\cite{step81} to
LiHoF$_{4}$. The transition temperature for the competing
ferromagnetic and spin-glass order parameters are given by the two
equations
\begin{eqnarray}
r_1&=&1-\sum_jx\tanh(J_{ij}/k_bT_c)=0 \\
r_2&=&1-\sum_jx\tanh^2(J_{ij}/k_bT_c)=0.
\end{eqnarray}
 For high temperatures $r_2 > r_1$ and ferromagnetic order persists, while,
depending on the lattice sums, $r_2$ may be smaller than $r_1$ for low
temperatures, in which case spin-glass ordering occurs. We have
evaluated the sums for the lattice appropriate for LiHoF$_{4}$ and found
that the solution favors spin-glass order for $x_c < 0.57$.

%\section{Discussion}
One reason for the discrepancy between the experimental results and
our calculations could lie in parts of the Hamiltonian that we have
neglected. The hyperfine coupling between nuclear and electronic spins
is important in the low-temperature regime and omitted in our
analysis. However, a recent study\cite{sche05} concluded that at zero
transverse field the hyperfine coupling would only renormalize the
Ising dipolar Hamiltonian and therefore it should not affect the phase
diagram qualitatively. In particular, it should not be a cause of the
spin-glass freezing. Another effect omitted in our simulation is
the generation of random magnetic fields due to the dilution, which
breaks the crystalline symmetry\cite{sche05, sche06,
  tabei:237203}. However, the effect of this term should be to
increase fluctuations and lower the critical temperature for both the
ferromagnetic and the spin-glass phase. It has even been argued that
off-diagonal dipolar terms destroy the spin glass transition at any
finite transverse field\cite{sche06}. We conclude that not only should
the omitted terms not cause a spin-glass transition, they also have
the potential of destroying the long-range glass order.

The analytic studies\cite{xu91, step81} yield the mean-field result
$T_c(x)\sim x$ in the limit of high dilution and therefore predict
long-range spin glass order extending all the way to $x=0$. This
result differs from both the experimental and our numerical studies,
which both predict a disordered system in the limit of extreme
dilution. It therefore appears that fluctuations not accounted for in
the theory are strong enough to cause a finite-dilution phase
transition at zero temperature. It would be of great interest to find
a theory that could account for the vanishing of the order in the
extreme dilution limit.

Numerical difficulties could also explain the difference between our
results and experiments. Glassy systems are notoriously hard to
equilibrate. Energy barriers between low-lying states cause
equilibration problems and make it hard to obtain reliable data for
large enough system sizes. The nearest-neighbor Ising spin glass has
been studied numerically for years, and only recently a consensus
seems to have developed concerning the glass transition. In our
simulations we see definite signs of equilibration problems at the
lowest temperatures. In particular we find that a decrease in $\langle
M^2\rangle$ as the temperature is lowered is a clear indicator that
the simulation does not reach equilibrium. However, having repeated
many of the simulations we believe that the data we show here is
reliable. The system sizes we consider (1000-4000 spins) are an order
of magnitude larger than in the previous study considering dipoles on
a cubic lattice\cite{snid05}, but we cannot entirely rule out that
finite-size effects are so strong in the high dilution limit that even
larger system sizes would be necessary to see the true thermodynamic
behavior of the model.

In order to resolve the differences it would also be important to have
more extensive experimental data. We are only aware of two
measurements\cite{rose90, rose93} of the spin glass transition in
LiHo$_{0.167}$Y$_{0.833}$F$_4$. In particular it would be of great
interest to have further data points in the region surrounding
$x=0.167$ to establish the extent and shape of the spin glass
phase. Further experimental data combined with more extensive Monte
Carlo simulation using parallel tempering, or other improved
equilibration techniques, should be able to resolve the present
differences.

\begin{acknowledgments}
  We acknowledge support by the G\"oran Gustafsson
  foundation. P.H. was supported by the Swedish Resarch Council.
  We are grateful to S. Girvin and A. Sandvik for helpful discussions.
\end{acknowledgments}

\bibliography{lihof}

\end{document}